\documentclass[11pt]{article}

\usepackage[final]{acl}

\usepackage{times}
\usepackage{latexsym}

\usepackage[T1]{fontenc}

\usepackage[utf8]{inputenc}

\usepackage{microtype}

\usepackage{graphicx}
\usepackage{booktabs}
\usepackage{multirow}

\usepackage[most]{tcolorbox}
\usepackage{pifont}  
\usepackage{amssymb} 

\usepackage{listings}
\definecolor{userblue}{RGB}{227, 242, 253}
\definecolor{userborder}{RGB}{66, 133, 244}
\definecolor{assistantgray}{RGB}{248, 249, 250}
\definecolor{assistantborder}{RGB}{108, 117, 125}
\definecolor{toolgreen}{RGB}{232, 245, 233}
\definecolor{toolborder}{RGB}{76, 175, 80}
\definecolor{errorred}{RGB}{255, 235, 238}
\definecolor{errorborder}{RGB}{220, 53, 69}
\definecolor{systempurple}{RGB}{243, 232, 255}
\definecolor{systemborder}{RGB}{128, 90, 213}

\newcommand{\usericon}{\ding{118}}        
\newcommand{\assistanticon}{\ding{72}}    
\newcommand{\toolicon}{\ding{115}}        
\newcommand{\resulticon}{\ding{51}}       
\newcommand{\erroricon}{\ding{54}}        
\newcommand{\systemicon}{\ding{108}}      

\newcommand{\stepmarker}[1]{%
  \vspace{4pt}%
  \noindent\textbf{\sffamily\small\textcolor{gray}{--- Step #1 ---}}%
  \vspace{2pt}%
}

\newtcolorbox{userbox}{
  enhanced,
  colback=userblue,
  colframe=userborder,
  boxrule=0.6pt,
  arc=2pt,
  left=4pt, right=4pt, top=2pt, bottom=2pt,
  fontupper=\small,
  fonttitle=\bfseries\footnotesize\sffamily,
  title={\usericon~User},
  coltitle=userborder,
  attach boxed title to top left={yshift=-1.5mm, xshift=3mm},
  boxed title style={colback=white, colframe=userborder, boxrule=0.4pt, arc=1.5pt, left=2pt, right=2pt, top=1pt, bottom=1pt}
}

\newtcolorbox{assistantbox}{
  enhanced,
  colback=assistantgray,
  colframe=assistantborder,
  boxrule=0.6pt,
  arc=2pt,
  left=4pt, right=4pt, top=2pt, bottom=2pt,
  fontupper=\small,
  fonttitle=\bfseries\footnotesize\sffamily,
  title={\assistanticon~Assistant},
  coltitle=assistantborder,
  attach boxed title to top left={yshift=-1.5mm, xshift=3mm},
  boxed title style={colback=white, colframe=assistantborder, boxrule=0.4pt, arc=1.5pt, left=2pt, right=2pt, top=1pt, bottom=1pt}
}

\newtcolorbox{toolbox}{
  enhanced,
  colback=toolgreen,
  colframe=toolborder,
  boxrule=0.6pt,
  arc=2pt,
  left=4pt, right=4pt, top=2pt, bottom=2pt,
  fontupper=\small\ttfamily,
  fonttitle=\bfseries\footnotesize\sffamily,
  title={\toolicon~Tool Call},
  coltitle=toolborder,
  attach boxed title to top left={yshift=-1.5mm, xshift=3mm},
  boxed title style={colback=white, colframe=toolborder, boxrule=0.4pt, arc=1.5pt, left=2pt, right=2pt, top=1pt, bottom=1pt}
}

\newtcolorbox{toolresultbox}{
  enhanced,
  colback=toolgreen!40,
  colframe=toolborder!80,
  boxrule=0.6pt,
  arc=2pt,
  left=4pt, right=4pt, top=2pt, bottom=2pt,
  fontupper=\small,
  fonttitle=\bfseries\footnotesize\sffamily,
  title={\resulticon~Result},
  coltitle=toolborder!80,
  attach boxed title to top left={yshift=-1.5mm, xshift=3mm},
  boxed title style={colback=white, colframe=toolborder!80, boxrule=0.4pt, arc=1.5pt, left=2pt, right=2pt, top=1pt, bottom=1pt}
}

\newtcolorbox{errorbox}{
  enhanced,
  colback=errorred,
  colframe=errorborder,
  boxrule=0.8pt,
  arc=2pt,
  left=4pt, right=4pt, top=2pt, bottom=2pt,
  fontupper=\small,
  fonttitle=\bfseries\footnotesize\sffamily,
  title={\erroricon~Failure Analysis},
  coltitle=errorborder,
  attach boxed title to top left={yshift=-1.5mm, xshift=3mm},
  boxed title style={colback=white, colframe=errorborder, boxrule=0.4pt, arc=1.5pt, left=2pt, right=2pt, top=1pt, bottom=1pt}
}

\newtcolorbox{systempromptbox}{
  enhanced,
  colback=systempurple,
  colframe=systemborder,
  boxrule=0.6pt,
  arc=2pt,
  left=4pt, right=4pt, top=2pt, bottom=2pt,
  fontupper=\small,
  fonttitle=\bfseries\footnotesize\sffamily,
  title={\systemicon~System Prompt},
  coltitle=systemborder,
  attach boxed title to top left={yshift=-1.5mm, xshift=3mm},
  boxed title style={colback=white, colframe=systemborder, boxrule=0.4pt, arc=1.5pt, left=2pt, right=2pt, top=1pt, bottom=1pt}
}

\newcommand{\errormark}[1]{\textcolor{errorborder}{\ding{54}~\textbf{#1}}}

\usepackage{amsmath,amsfonts,bm}

\def\eqref#1{equation~\ref{#1}}

\def\1{\bm{1}}

\DeclareMathAlphabet{\mathsfit}{\encodingdefault}{\sfdefault}{m}{sl}
\SetMathAlphabet{\mathsfit}{bold}{\encodingdefault}{\sfdefault}{bx}{n}

\title{Breaking the Evaluation Paradox: Evaluating High-Entropy Search with Computationally Irreducible Constraints}

\author{
Juntao Wu\textsuperscript{1,2}\thanks{\ Equal contribution. Work done during internship at Tencent.},
Wei Wen\textsuperscript{2},
Xianting Huang\textsuperscript{3},
Shuai Pang\textsuperscript{1},
Ruizhi Qiao\textsuperscript{2},\\
\textbf{Xing Sun}\textsuperscript{2},
\textbf{Ke Wang}\textsuperscript{1}\thanks{\ Corresponding author.}\\
\textsuperscript{1}Jinan University \quad
\textsuperscript{2}Tencent Youtu Lab \quad
\textsuperscript{3}The Chinese University of Hong Kong\\
\texttt{\{jtitor, wangke\}@jnu.edu.cn} \quad
\texttt{ps2023104429@stu2023.jnu.edu.cn}\\
\texttt{\{jawnrwen, ruizhiqiao, winfredsun\}@tencent.com} \quad
\texttt{alisonwww@link.cuhk.edu.hk}
}

\begin{document}
\maketitle

\begin{abstract}
Evaluating the exhaustive search capabilities of large language models (LLMs) is plagued by a fundamental paradox: 
verifying completeness requires complete ground truth, yet high-entropy enumeration tasks make such ground truth impossible for humans to create. This causes benchmarks to systematically penalize models for outperforming their human annotators. Despite rapid progress in web-search and deep research agents---which now issue hundreds of queries, traverse diverse sites, and synthesize long reports---evaluation still largely relies on partially annotated answer sets, LLM-based judges, or single-answer questions that avoid genuinely exhaustive search scenarios. 

We break this paradox by shifting the evaluation paradigm from simulating a messy reality to constructing computationally pure challenges. We introduce VERITAS (Verifiable Traversal Assessment for Search), a framework built on the principle of \textbf{computationally irreducible constraints}. By introducing novel, non-optimizable constraints, we create verifiable, sparse-answer search tasks that are computationally equivalent to exhaustive enumeration. These constraints are easy to verify but impossible for LLMs or search engines to optimize, forcing agents to genuinely traverse the entire search space. VERITAS can automatically generate a virtually infinite number of test cases with perfect ground truth and precise difficulty control, with marginal instance cost dominated by hash computations. This provides not only a robust benchmark for evaluating systematic exploration under uncertainty but also a scalable method for generating training data to improve these crucial, yet underdeveloped, capabilities.
\end{abstract}

\section{Introduction}

Large language models (LLMs) achieve strong performance on factual question-answering benchmarks with single, deterministic targets~\citep{kwiatkowski2019natural,joshi2017triviaqa}, but their behavior on high-entropy queries (HEQs)---tasks that require enumerating large open-ended answer sets---remains underexplored. Many real-world information needs resemble HEQs: success depends on recall over a broad candidate space rather than on one ``best'' answer, and agents must decide when additional exploration is no longer worthwhile.

Recent web-search and ``deep research'' agents push LLMs directly into this regime: commercial and open-source systems routinely issue hundreds of queries, traverse diverse sites, and synthesize long reports from heterogeneous evidence~\citep{openaidr,perplexity,bc_en,tao2507webshaper,li2025webthinker}. Benchmarks such as GAIA and BrowseComp~\citep{mialon2023gaia,bc_en,bc_zh,wei2504browsecomp} evaluate multi-step browsing on ``hard to find but easy to verify'' questions, while WideSearch and DeepWideSearch~\citep{widesearch,deepwidesearch} target breadth and depth of coverage over thousands of entities. With the growing attention to complex information needs in real-world scenarios, however, the majority of these benchmarks rely on partially annotated answer pools, which not only incur considerable costs but also risk penalizing agents that discover valid items beyond the pool as hallucinating.

We call this the \emph{evaluation paradox} for high-entropy search: exhaustive search should be judged against a complete ground-truth set, yet for HEQs such ground truth is practically intractable to obtain. The same open-endedness that makes these tasks hard for models prevents annotators from certifying completeness, so annotated answer pools are almost surely incomplete. Under standard evaluation protocols, any correct item outside the annotated set is indistinguishable from an error, and systems that discover more valid items can be penalized as hallucinating. As a result, measured performance is effectively bounded by annotation coverage rather than by an agent's true search capability, even when wide-search benchmarks attempt to increase coverage with large manually curated pools~\citep{widesearch,deepwidesearch}.

In this work we take a complementary approach. Instead of approximating real-world HEQs with partially annotated answer sets, we construct \emph{synthetic objectives} on top of real web-scale domains. Our VERITAS benchmark uses cryptographic hash constraints to turn high-entropy enumeration into sparse-answer search tasks that are \emph{easy to verify but impossible to optimize semantically}. Agents still have to traverse genuine web content (e.g., artist discographies, model hubs) using the same kinds of search and browsing tools as in existing benchmarks, but success reduces to finding items whose identifiers satisfy a non-optimizable hash constraint, enabling exact labels and scalable instance generation.

\begin{table*}[t]
\centering
\small
\begin{tabular}{lccccc}
\toprule
\textbf{Benchmark} & \textbf{Wide} & \textbf{Scalable} & \textbf{Human} & \textbf{Verification} & \textbf{Core Issue} \\
 & \textbf{Degree} & & \textbf{Depend} & \textbf{Difficulty} & \\
\midrule
GAIA & $\star$ & No & Yes & Easy & Not Wide \\
 & Low & Hard & Expert & Single answer & Human \\
 & (Single ans) & & & Easy verification & dependent \\
\midrule
BrowseComp & $\star$ & No & Yes & Easy & Not Wide \\
(OpenAI 2025) & Low & Hard & Full & Single answer & Human \\
 & (Single ans) & & & Easy verification & dependent \\
\midrule
WideSearch & $\star\star\star$ & No & Yes & \textbf{Hard} & Human annotation \\
 & Medium & Hard & Full & Many entities & May punish \\
 & (Wide) & & & LLM unstable & better answers \\
\midrule

\textbf{VERITAS} & \textbf{$\star\star\star\star\star$} & \textbf{Yes} & \textbf{No} & \textbf{Easy} & \\
\textbf{(Ours)} & \textbf{Extreme} & \textbf{Easy} & \textbf{Zero} & \textbf{Easy} & \\
 & \textbf{(Unlimited)} & & & \textbf{verification} & \\
\bottomrule
\end{tabular}
\caption{Comparison of VERITAS with existing web-search benchmarks. ``Wide degree'' indicates breadth of search space (stars). VERITAS uniquely combines extreme breadth, exact automated verification, and no human dependence, whereas prior benchmarks trade off coverage against verification difficulty.}
\label{tab:benchmark_comparison}
\end{table*}

Our contributions are threefold:
\begin{enumerate}
     \item We formalize the evaluation paradox for high-entropy search tasks and analyze why it is difficult to resolve within annotation-based paradigms.
     \item We introduce VERITAS, a benchmark that uses computationally irreducible, hash-based constraints to construct scalable, automatically verifiable tests of exhaustive search capabilities without human labels or LLM judges.
     \item We combine theoretical and empirical analyses to show that hash-constrained sparse search is computationally equivalent to exhaustive enumeration, and we evaluate several state-of-the-art web agents on VERITAS to characterize their failure modes on high-entropy search tasks.
\end{enumerate}

\section{Related Work}

Evaluation of information-seeking systems has evolved from static QA benchmarks such as HotpotQA and related multi-hop datasets~\citep{yang2018hotpotqa,ho2020constructing,trivedi2022musique,press2022measuring} to dynamic, agent-centric web benchmarks that require navigation and tool use~\citep{mialon2023gaia,bc_en,bc_zh,wei2504browsecomp,widesearch,deepwidesearch}. WebArena~\citep{zhou2023webarena} introduced self-hosted realistic web environments for autonomous agents, establishing foundational evaluation protocols. Subsequently, AgentBench~\citep{liu2023agentbench} evaluates LLM agents in 8 environments, including web browsing. Specifically, GAIA and BrowseComp construct ``hard to find but easy to verify'' tasks that emphasize multi-step browsing under fuzzy, underspecified instructions, whereas WideSearch and DeepWideSearch explicitly target breadth and depth of search over thousands of entities. All of these benchmarks ultimately rely on large but still incomplete manually annotated answer pools, so coverage limitations remain a central bottleneck.

Commercial ``deep research'' products such as those from OpenAI, Perplexity, Gemini, Kimi, Doubao, and Grok~\citep{openaidr,perplexity,geminidr,kimidr,doubaodr,grokdr}, together with open-source frameworks like WebShaper, WebSailor, WebThinker and WebDancer~\citep{tao2507webshaper,li2025websailor, li2025webthinker, wu2025webdancer}, push LLM agents into long-horizon web settings and typically evaluate them on GAIA, BrowseComp, WideSearch, and related suites~\citep{zhang2025fromwebsearch,xi2025surveydeepsearch}. Alongside these explorations and the application of reinforcement learning to search agents~\citep{song2025r1,jin2025search}, VERITAS's potential as an important data source correspondingly emerges.

 Moreover, recent surveys on LLM-based agents and deep research~\citep{guo2024large,wang2024survey,zhang2025fromwebsearch,xi2025surveydeepsearch} likewise highlight that high-entropy, partially labeled conditions make it difficult to measure exhaustive search capability. VERITAS is complementary to these efforts: by introducing \textbf{computationally irreducible constraints} via cryptographic hash functions, it isolates the enumeration component of deep research and provides exact, automatically scalable supervision rather than relying on manually constructed answer sets.

\section{Method}

\subsection{Evaluation Paradox}

\textbf{Exhaustive search tasks are fundamentally ill-posed for evaluation: verifying completeness requires the very ground truth that the task itself makes intractable to obtain.}

The difficulty stems from two inherent properties of high-entropy queries. First, relevant information is \textbf{severely fragmented} across many loosely connected web sources, with no single canonical page from which to harvest complete answer sets. Second, such tasks provide \textbf{no principled stopping rule}: neither annotators nor agents can know when further search will yield additional items. This combination makes literal completeness unattainable at web scale, so any human-constructed ground truth is almost guaranteed to be incomplete---and evaluation protocols may inadvertently penalize models that exceed human coverage while rewarding those that simply reproduce human omissions.

Consider the query: ``Has the virtual singer KAF ever hosted a radio show, and if so, which show has the most episodes?'' In a pilot analysis, one annotator concluded KAF has never hosted any show; another identified a short series with few episodes; a more systematic audit later found multiple series with substantially more episodes. This discrepancy reflects not human error, but the inherent difficulty---annotators face the same fragmentation and stopping challenges that define the task itself.

Addressing this paradox requires reformulating the evaluation problem, rather than attempting to obtain ever more exhaustive annotations.

\begin{figure*}[t]
  \centering
  \includegraphics[width=0.95\textwidth]{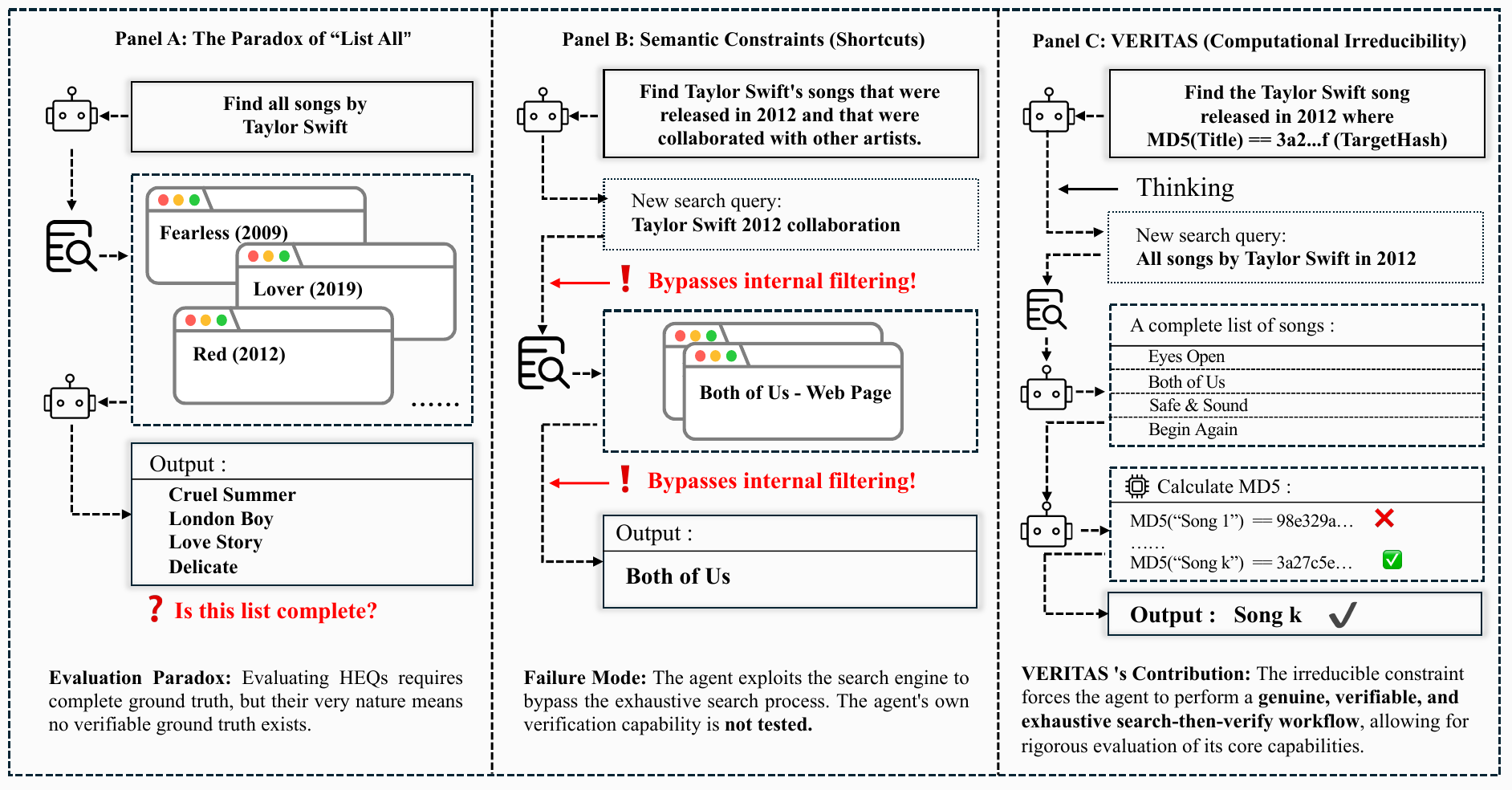}
  \caption{Evaluation paradox for high-entropy queries and our reformulation. (a) Exhaustive annotation is intractable. (b) Naively strengthening semantic constraints does not resolve the coverage problem. (c) Reformulating the task around automatically verifiable, non-optimizable constraints preserves exhaustive-search difficulty while enabling exact verification.}
  \label{fig:evaluation_paradox}
\end{figure*}

\subsection{Computational Equivalence}

\textbf{The ill-posedness of ``list all~$X$'' benchmarks stems from annotation limitations, not evaluation goals; any task that forces the same high-entropy search without exploitable shortcuts is computationally equivalent.}

From an evaluation perspective, the intent is not literally to obtain exhaustive lists, but to probe two deeper capabilities. First, an effective agent should carry out \textbf{wide and systematic search} over a large, loosely structured space, so that, given sufficient budget, the probability of missing salient items is small. Second, it should maintain \textbf{calibrated residual uncertainty}, distinguishing between ``the space is probably covered'' and ``further exploration is still warranted'' when deciding whether to stop. In this view, ``list all~$X$'' is merely one syntactic vehicle for these desiderata. Our goal is to decouple the underlying search-and-coverage challenge from the brittle requirement of exact enumeration.

A simple analogy clarifies this equivalence: searching for the Ace of Spades in a well-shuffled deck. Whether we ask to ``list all 52 cards'' or to ``find the Ace of Spades,'' the worst-case search cost is the same---one must inspect cards sequentially until the target appears, requiring $\Theta(N)$ operations over an unstructured space of size $N$. The identity of the target provides no algorithmic shortcut; the difficulty arises from traversing an unstructured space.

We apply the same idea to high-entropy web queries. Instead of ``List all songs by artist X,'' we ask ``Find the unique song by artist X whose title satisfies property $Y$.'' When property $Y$ is \textbf{non-optimizable}---offering no usable signal to retrieval mechanisms---the agent must approximate the original $O(N)$ exhaustive search: gather candidates and check each against $Y$. By choosing $Y$ to be \emph{verifiable but unoptimizable}, we force agents to engage with the same high-entropy search problem while allowing evaluation to depend on a single, easily checked answer rather than a fragile exhaustive list.

Concretely, our benchmark instantiates $Y$ using automatically verifiable yet non-optimizable constraints on item identifiers or derived attributes, so that success reduces to finding a small number of constraint-satisfying items in a large unstructured space. Figure~\ref{fig:evaluation_paradox} provides a high-level schematic of this shift from ill-posed exhaustive annotation to a constrained, verifiable search task. The next subsection formalizes this abstract model and analyzes its scaling behavior.

\subsection{Asymptotic Analysis}

We now formalize why finding $k$ sparse constraint-satisfying items under a non-optimizable condition is computationally comparable to full enumeration. This provides a tractable model for interpreting our benchmark results.

\textbf{Problem setup}: Given $N$ items in an unstructured search space, $k \leq N$ items satisfy a non-optimizable constraint. An agent with search budget $T$ must find \textbf{all} $k$ matching items. Success requires $k$ correct items; partial matches count as failure.

\textbf{Expected search cost}: Assume the $k$ target items are uniformly distributed among $N$ positions. Under systematic or random search without replacement, the expected position of the last (k-th) discovered target is:
\begin{equation}
\mathbb{E}[L_k] = \frac{(N+1) \cdot k}{k+1}.
\end{equation}

This is the standard order-statistics result: if the $k$ targets occupy $k$ random positions among $N$ items, the expected position of the last one is $(N+1) \cdot k/(k+1)$.

\textbf{Asymptotic behavior}: For any fixed $k \geq 1$, $\mathbb{E}[L_k] = \Theta(N)$. Moreover,
\begin{align}
\text{Relative cost: } \frac{\mathbb{E}[L_k]}{N} &= \frac{k}{k+1} \xrightarrow{k \to \infty} 1.
\end{align}

Even for small $k$, the expected search cost is substantial: $\mathbb{E}[L_1] = (N+1)/2 \approx N/2$, $\mathbb{E}[L_2] = 2(N+1)/3 \approx 2N/3$, $\mathbb{E}[L_4] = 4(N+1)/5 \approx 4N/5$. Finding even a single sparse item therefore requires searching about half the space on average.

\textbf{Success probability under budget constraints}: Given finite search budget $T < N$, the probability of finding all $k$ targets follows a hypergeometric distribution. If the agent searches exactly $T$ items uniformly at random,
\begin{equation}
P_{\text{success}}(k, T, N) = \frac{\binom{N-k}{T-k}}{\binom{N}{T}} = \prod_{j=0}^{k-1} \frac{T-j}{N-j}.
\end{equation}

For large $N$ with small $k$ and $T \ll N$, this probability can be approximated by
\begin{equation}
P_{\text{success}}(k, T, N) \approx \left(\frac{T}{N}\right)^k.
\end{equation}

This expression highlights a sharp dependence on $k$: as $k$ increases, success probability decreases roughly exponentially unless $T$ is on the order of $N$. Specifically:
\begin{itemize}
\item $k=1$: $P \approx T/N$ (linear dependence)
\item $k=2$: $P \approx (T/N)^2$ (quadratic drop)
\item $k=4$: $P \approx (T/N)^4$ (quartic drop)
\item $k=N$: $P = 1$ iff $T = N$ (deterministic threshold).
\end{itemize}

\textbf{Controlling difficulty via search space size}: The success probability formula $P_{\text{success}}(k, T, N) \approx (T/N)^k$ reveals two independent mechanisms for controlling task difficulty: adjusting $k$ (number of targets) or $N$ (search space size). While the previous analysis focused on varying $k$, our benchmark design also exploits $N$ as a difficulty lever.

Consider our Medium-tier tasks, which require searching across 7 independent sources (e.g., finding which of 7 artists has a song satisfying the target constraint). If each artist has approximately $n$ songs, the effective search space becomes $N = 7n$, substantially larger than a single-artist search with $N = n$. For fixed $k=1$ and search budget $T$, the success probability scales as:
\begin{equation}
\frac{P_{\text{single}}}{P_{\text{medium}}} = \frac{T/n}{T/(7n)} = 7.
\end{equation}

This $7\times$ difficulty multiplier arises purely from increasing $N$, demonstrating that enlarging the search space (the denominator in $(T/N)^k$) is as effective as increasing the numerator $k$ for creating challenging tasks. The Hard and Extra Hard tiers further increase $N$ through deep attribute extraction and nested task hierarchies, respectively, creating progressively larger search spaces while maintaining irreducibility under the chosen non-optimizable constraints.

Taken together, these observations show that our transformation—from ``enumerate all'' to ``find $k$ constraint-satisfying items''—preserves the $O(N)$ computational irreducibility of the task while enabling deterministic verification. The non-optimizable constraint removes optimization shortcuts without altering the fundamental search complexity. Moreover, the $(T/N)^k$ framework provides principled control over difficulty through both $k$ and $N$, enabling fine-grained benchmark design.

We now turn to how to construct constraints that realize this irreducible search behavior in practice.

\begin{figure*}[t]
  \centering
  \includegraphics[width=0.98\textwidth]{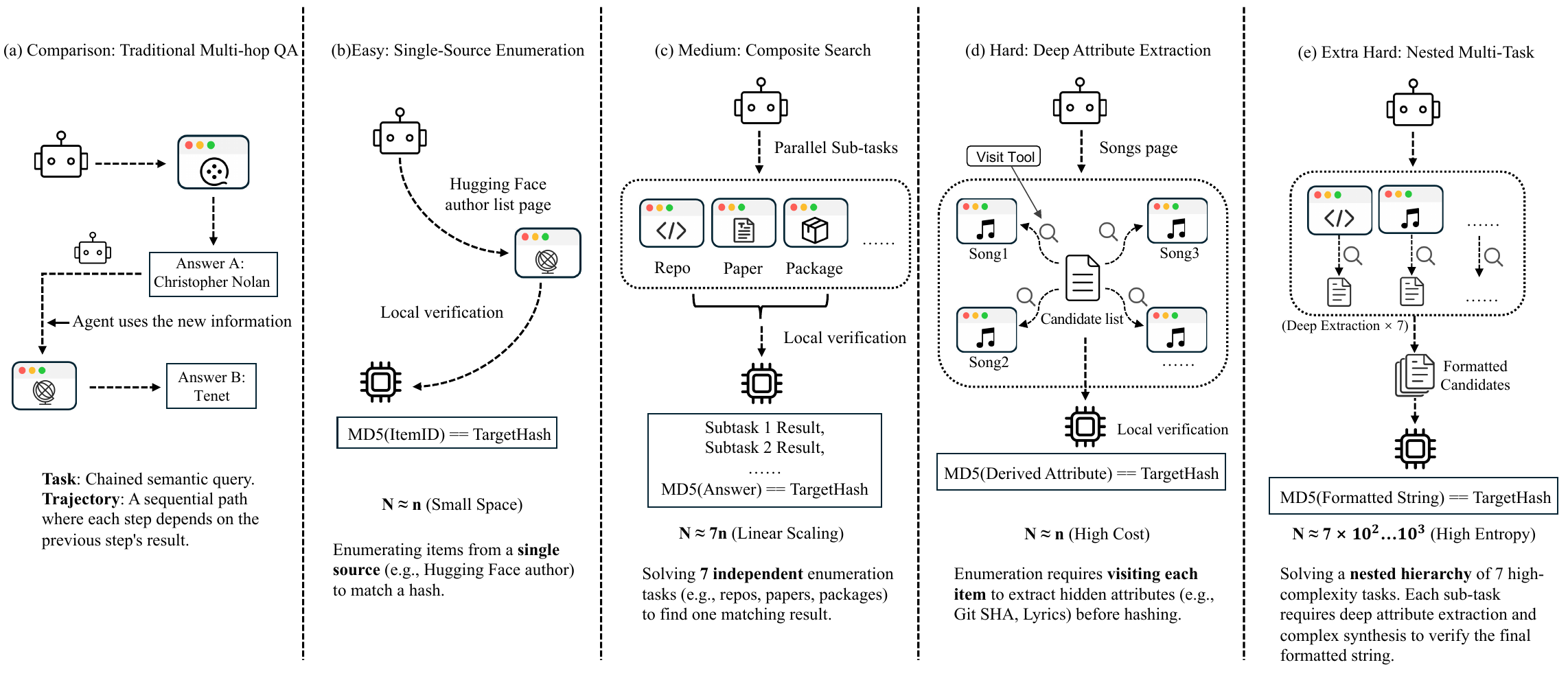}
  \caption{Progressive difficulty tiers in VERITAS. Higher tiers enlarge the search space and computational burden while preserving hash-based irreducibility.}
  \label{fig:data}
\end{figure*}

\subsection{Semantic Constraints}

\textbf{Semantic constraints fail as non-optimizable barriers: modern systems can delegate them to search engines, collapsing the intended $O(N)$ exhaustive search into $O(1)$ lookup.}

The most direct response to the evaluation paradox is to introduce additional semantic constraints on target items. For instance, we might ask: ``Find a Jay Chou song released in 2003, with a ballad style, lasting over four minutes, and featuring piano.'' At face value, this appears to require enumerating candidates and verifying each. In practice, however, indexing and filtering mechanisms explicitly exploit semantic information, so temporal, categorical, or relational constraints become \textbf{signals for optimization} rather than obstacles. The model can submit the entire constraint set as a query and retrieve a small candidate pool directly.

More broadly, many ``fuzzy'' web tasks decompose into two stages: first \emph{enumerate} a broad candidate space, then \emph{filter} using semantic conditions. The intrinsic difficulty lies in enumeration; once candidates are available, filtering is straightforward. Our benchmark therefore targets enumeration directly. To recover computational equivalence with high-entropy queries while avoiding semantic shortcuts, we require constraints that are \textbf{verifiable but unoptimizable}: easily checked once a candidate is known, yet providing no useful guidance to search engines.

\subsection{Cryptographic Hashes}

\textbf{Cryptographic hash functions provide ideal non-optimizable constraints through two properties: one-way verification and semantic nullity.}

\textbf{One-way verification}: Pre-image resistance makes reversing the hash computationally infeasible, yet verification is trivial---hash the candidate and compare. This is a lock that only opens when you already have the key.

\textbf{Semantic nullity}: Hash outputs bear no exploitable relationship to the semantic content of their inputs. The avalanche effect ensures that small input changes produce seemingly unrelated outputs, rendering the hash useless as a search query. Search engines cannot leverage it to narrow the search space.

Under these conditions, solving the task requires approximate $O(N)$ exhaustive search. Models cannot use the hash value to guide retrieval; they must enumerate candidate items and verify each one. This transformation preserves the underlying computational challenge while enabling reliable, scalable verification. A hash-constrained task with $N$ implicit candidates and $k$ true pre-images can be modeled as selecting $k$ targets uniformly at random from an unstructured space of size $N$, matching the abstract setting analyzed previously.

\subsection{VERITAS Tasks}

\textbf{VERITAS implements four progressive difficulty tiers that systematically stress-test exhaustive search under increasing computational irreducibility} (Figure~\ref{fig:data}).

Each tier enlarges the effective search space $N$ or increases task complexity while preserving hash-based irreducibility. This provides principled difficulty control: per the $(T/N)^k$ framework, both expanding $N$ and increasing structural complexity reduce success probability for budget-constrained agents.

\textbf{Easy (Single Source)}: Enumerate all items from a single source (e.g., ``all models by author \texttt{google} on HuggingFace'') and identify the item whose identifier's MD5 hash matches the target. Assesses basic enumeration discipline.

\textbf{Medium (Composite Search)}: Seven independent enumeration subtasks; the agent must solve each, then determine which result produces the target MD5 hash. Evaluates coordination of multi-stage exhaustive search.

\textbf{Hard (Deep Attribute Extraction)}: Enumerate items from a source and identify which item's \textbf{derived attribute} (e.g., a Git commit SHA) yields the target hash. Emphasizes deep attribute extraction from distributed sources.

\textbf{Extra Hard (Nested Multi-Task)}: Seven high-complexity enumeration tasks; construct complete formatted answers (e.g., ``RepoName~$\rightarrow$~SHA'') and determine which formatted string has the target hash. Probes sustained search across deeply nested task hierarchies.

\section{Experiments}

\begin{table*}[!t]
\centering
\small
\caption{Performance on VERITAS benchmark across difficulty tiers. We report Pass@4 success rate (\%) and the average number of tool calls per attempt, split into Search and Visit. Best results are highlighted in \textbf{bold}.}
\label{tab:main_results}
\begingroup
\setlength{\tabcolsep}{2.0pt}
\renewcommand{\arraystretch}{1.15}
\begin{tabular*}{\textwidth}{@{\extracolsep{\fill}}lrrrrrrrrrrrr}
\toprule
\multirow{3}{*}{\textbf{Model}} & \multicolumn{4}{c}{\textbf{Medium}} & \multicolumn{4}{c}{\textbf{Hard}} & \multicolumn{4}{c}{\textbf{Ex-Hard}} \\
\cmidrule(lr){2-5} \cmidrule(lr){6-9} \cmidrule(lr){10-13}
& \multicolumn{2}{c}{\textbf{Success}} & \multicolumn{2}{c}{\textbf{Tool Calls}} & \multicolumn{2}{c}{\textbf{Success}} & \multicolumn{2}{c}{\textbf{Tool Calls}} & \multicolumn{2}{c}{\textbf{Success}} & \multicolumn{2}{c}{\textbf{Tool Calls}} \\
\cmidrule(lr){2-3} \cmidrule(lr){4-5} \cmidrule(lr){6-7} \cmidrule(lr){8-9} \cmidrule(lr){10-11} \cmidrule(lr){12-13}
& \textbf{Pass@4} & \textbf{Avg@4} & \textbf{Search} & \textbf{Visit} & \textbf{Pass@4} & \textbf{Avg@4} & \textbf{Search} & \textbf{Visit} & \textbf{Pass@4} & \textbf{Avg@4} & \textbf{Search} & \textbf{Visit} \\
\midrule
\multicolumn{13}{c}{\textit{Open-Source Models}} \\
\addlinespace[0.25ex]
GLM-4 & \textbf{17.9} & 5.7 & 23.4 & 10.1 & 12.2 & 5.6 & 14.9 & 8.8 & 3.3 & 2.5 & 23.2 & 12.2 \\
MiniMax-M2 & 12.5 & 4.4 & 14.9 & 8.7 & 6.1 & 1.5 & 17.4 & 7.8 & 3.4 & 0.8 & 17.7 & 10.0 \\
DeepSeek-v3.1 & 10.3 & 3.8 & 15.5 & 8.8 & \textbf{14.3} & 7.1 & 9.2 & 4.1 & 3.4 & 0.9 & 16.0 & 9.2 \\
Qwen3-235B-A22 & 5.3 & 2.6 & 16.8 & 13.1 & 6.2 & 2.6 & 8.0 & 3.9 & \textbf{3.6} & 2.6 & 16.9 & 12.4 \\
Kimi-K2 & 2.5 & 0.6 & 16.2 & 8.5 & 10.2 & 2.6 & 18.0 & 6.5 & 0.0 & 0.0 & 26.9 & 9.5 \\
\addlinespace[0.5ex]
\midrule
\multicolumn{13}{c}{\textit{Closed-Source Models}} \\
\addlinespace[0.25ex]
GPT-4o & 6.0 & 1.5 & 5.3 & 4.9 & 2.4 & 2.4 & 2.2 & 2.1 & 0.0 & 0.0 & 5.1 & 4.7 \\
Gemini-2.5-Pro & 12.5 & 6.2 & 14.8 & 9.9 & 8.2 & 3.6 & 9.6 & 4.4 & 3.3 & 1.7 & 14.4 & 10.8 \\
Gemini-3-Pro & 30.0 & 18.8 & 84.7 & 6.8 & \textbf{36.7} & 21.9 & 92.5 & 6.8 & 6.7 & 3.3 & 125.8 & 9.3 \\
GPT-5 & \textbf{47.5} & 27.5 & 7.8 & 101.8 & 32.7 & 25.5 & 7.3 & 16.4 & \textbf{20.0} & 20.0 & 9.3 & 27.8 \\
\bottomrule
\end{tabular*}
\endgroup
\end{table*}

\begin{figure*}[!t]
    \centering
    \includegraphics[width=\textwidth]{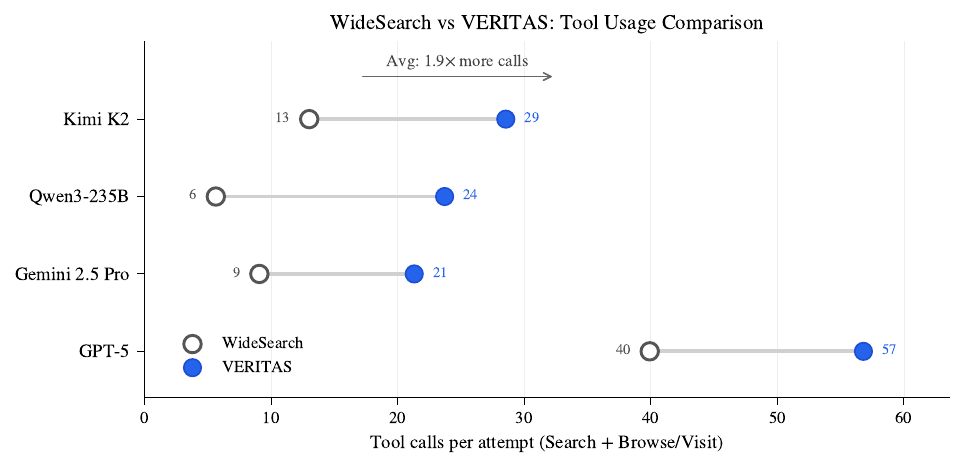}
    \caption{Tool calls per attempt for models evaluated on both WideSearch and VERITAS. VERITAS consistently elicits $1.4\times$--$4.2\times$ more tool calls than WideSearch, yet success rates remain low (Table~\ref{tab:main_results}), confirming that VERITAS removes semantic shortcuts and forces genuine exhaustive enumeration.}
    \label{fig:toolcalls_shared}
\end{figure*}

\textbf{VERITAS reveals a clear capability hierarchy: frontier models (GPT-5, Gemini-3-Pro) substantially outperform others, yet even they struggle on Extra Hard, with success rates dropping by $2$--$5\times$.} We evaluate nine models across Medium, Hard, and Extra Hard tiers to characterize both capability gaps and failure modes under computationally irreducible constraints.

\subsection{Agent Tools}

\textbf{Agents are equipped with four tools---Search, Visit, Exec Python, and Answer---providing capabilities comparable to human web researchers.}

\textbf{Search:} Provides access to Google search, returning top-10 results (title, snippet, URL) for each query. Supports multiple simultaneous queries.

\textbf{Visit:} Accesses specific web pages with dedicated extraction goals. Following WebSailor's approach, we use Qwen3-235B-A22 to extract goal-relevant information from each page.

\textbf{Exec Python:} Executes Python code for computations, including MD5 hash verification of candidate items.

\textbf{Answer:} Submits the final solution for evaluation.

\subsection{Experimental Setup}

\textbf{We evaluate across Medium/Hard/Extra Hard tiers on balanced multilingual queries, with strict all-or-nothing success criteria.}

We deliberately exclude Easy tier to focus on realistic high-entropy challenges. The dataset maintains balanced distribution across English, Chinese, and Japanese, ensuring robust cross-lingual evaluation.

We evaluate nine models spanning diverse architectures and capability levels: \textbf{GLM-4}, \textbf{MiniMax-M2}, \textbf{DeepSeek-v3.1}, \textbf{Qwen3-235B-A22}, \textbf{Kimi-K2}, \textbf{GPT-4o}, \textbf{Gemini-2.5-Pro}, and the latest frontier models \textbf{GPT-5} and \textbf{Gemini-3-Pro}. Success requires finding \emph{all} items matching the target MD5 hash---partial matches count as failure.

We report \textbf{Pass@4} (success rate across 4 attempts) and track tool usage via average \textbf{Search} and \textbf{Visit} calls per attempt.

\subsection{Main Results}

\textbf{VERITAS effectively discriminates model capabilities: frontier models (GPT-5, Gemini-3-Pro) substantially outperform others, yet even they see dramatic performance collapse on Extra Hard.}

Table~\ref{tab:main_results} reveals a clear \textbf{capability hierarchy}. GPT-5 achieves 47.5\% Pass@4 on Medium and 32.7\% on Hard---far surpassing other models. Gemini-3-Pro follows with 30.0\% and 36.7\% respectively. In contrast, the remaining seven models cluster between 2.5--17.9\% on Medium and 2.4--14.3\% on Hard. This stark separation demonstrates that VERITAS provides meaningful signal for distinguishing frontier capabilities from current-generation models.

Crucially, \textbf{Extra Hard remains challenging even for frontier models}. GPT-5 drops from 47.5\% (Medium) to 20.0\% (Extra Hard)---a $2.4\times$ reduction. Gemini-3-Pro drops more sharply from 36.7\% (Hard) to just 6.7\% (Extra Hard). Meanwhile, most other models collapse to 0--3.6\% on Extra Hard, with GPT-4o and Kimi-K2 achieving exactly 0\%. This pattern confirms that Extra Hard tier probes a genuine capability frontier: it separates GPT-5 (the only model maintaining double-digit performance) from all others, while still exposing substantial room for improvement.

The \textbf{difficulty gradient works as designed}: success rates drop consistently from Medium to Hard to Extra Hard across all models. This validates our $(T/N)^k$ theoretical framework---as the effective search space $N$ expands through composite search, deep attribute extraction, and nested hierarchies, success probability decreases predictably.

Finally, the gap between Pass@4 and Avg@4 reveals search strategy quality. GPT-5 shows notably higher consistency (Avg@4 of 27.5\%, 25.5\%, 20.0\% across tiers) compared to other models where Avg@4 often falls far below Pass@4, indicating that weaker models occasionally stumble upon solutions but lack systematic search strategies.

\subsection{Scaling under Hash Constraints}

\textbf{Models invest substantial search effort (20--30+ tool calls per query) yet achieve $<$12\% success, confirming that VERITAS forces genuine exhaustive enumeration without semantic shortcuts.}

Tool-usage statistics reveal a striking effort-outcome mismatch. GLM-4 issues on average 31.3 tool calls per query (19.8 Search, 10.0 Visit), Kimi-K2 28.5, and DeepSeek-v3.1 21.9---far exceeding typical single-answer QA benchmarks such as GAIA (average $<$10 calls). Despite examining dozens of web pages, success rates remain below 12\%. Models recognize the need for exhaustive exploration but fail to achieve sufficient coverage.

Figure~\ref{fig:toolcalls_shared} provides direct evidence: for the same underlying models, VERITAS elicits $1.4\times$--$4.2\times$ more tool calls than WideSearch, yet Pass@4 remains low. This gap underscores that VERITAS removes semantic shortcuts and forces agents into genuine enumeration regimes.

This pattern precisely matches the $(T/N)^k$ scaling analysis: as effective search space $N$ grows from Medium to Extra Hard while tool budgets remain fixed, success probability collapses across all models. VERITAS thus serves as a sharper stress test of exhaustive-search capability rather than surface-level browsing.




\section{Conclusion}

We argued that the core difficulty in modern web search and ``deep research'' is not the length of reasoning chains, but the \emph{entropy} of the search space: the need to enumerate large, weakly structured candidate sets and to know when it is safe to stop. Through theoretical analysis, we showed that finding a few sparse, hash-constrained targets is computationally equivalent to exhaustive enumeration, providing a path to automatically verifiable evaluation of this capability.

Building on this analysis, VERITAS instantiates cryptographic hash constraints over realistic, web-scale high-entropy queries to create tasks where any successful agent must effectively ``list everything that could possibly match'' before applying a cheap verification step. Experiments across strong web agents reveal that, despite heavy tool use, models still collapse on these tasks as search spaces grow, highlighting a fundamental gap between current systems and truly exhaustive search.

Looking forward, the same properties that make VERITAS attractive as a benchmark---infinite automatic generation, exact labels, and tunable difficulty---also make it a promising source of training data for future search agents, for example via reinforcement learning or RL-style post-training. We view our work as a first step towards systematically teaching models not just to find good answers, but to approach the harder goal of finding \emph{all} relevant answers under uncertainty.

\section*{Limitations}

VERITAS currently serves purely as an evaluation benchmark. Although its automatic, verifiable structure makes it well suited for reinforcement learning or other post-training methods, we have not yet applied such training to models using VERITAS-generated data; exploring this direction is an important target for future work.

\section*{Acknowledgements}
This work was by National Key Program of National Natural Science of China (Grant No.82430108).

\bibliography{anthology}

\newpage
\clearpage
\appendix

\appendix

\section{Theoretical Insights: The Epistemology of Search Difficulty}
\label{sec:appendix_insights}

Due to space limitations in the main text, which focuses primarily on our methodology and experimental results, we were unable to fully detail our theoretical analysis of search difficulty. In this section, we share our deconstruction of ``difficulty'' in agent search tasks, distinguishing between logical depth and search entropy, and explaining why existing benchmarks often fail to measure true exhaustive search capabilities.

\subsection{The Unreliability of Depth as a Proxy: Multi-hop $\neq$ Difficulty}

\textbf{Multi-hop structure is an unreliable proxy for difficulty because it conflates logical depth with search entropy; a 10-hop chain with branching factor 1 is computationally trivial, while a 2-hop query over an unbounded space may be intractable.}

A long-standing view in the community is that ``multi-hop questions'' intrinsically represent high difficulty because agents must reason across multiple nodes. We argue that using multi-hop structure as a direct proxy for difficulty is unreliable because it confuses \textbf{logical depth} with \textbf{search entropy}. The distinction is critical: depth counts steps, but entropy measures the size of the search space at each step. A deep chain through narrow corridors is easy; a shallow query into an ocean of possibilities is hard.

\subsubsection{Low Entropy in Deterministic Chains}

\textbf{When relationships are deterministic, multi-hop merely increases operations linearly---the computational burden is $O(k)$, not exponential.} Think of it as walking a straight hallway versus navigating a branching maze: the hallway may be long, but you never face a choice.

Consider the following classic multi-hop question:
\begin{quote}
    ``What is the profession of the grandfather of the husband of the first female Nobel Prize winner?''
\end{quote}
Although this problem has a structure of $\ge 4$ hops, it is not computationally difficult because the branching factor at each step is low (often one-to-one). The search path is deterministic:
\begin{multline*}
\text{Find(Nobel\_Women)} \xrightarrow{1:1} \text{Marie Curie} \\
\xrightarrow{1:1} \text{Pierre Curie} \xrightarrow{1:1} \text{Grandfather}
\end{multline*}
When relationships are deterministic, multi-hop merely increases the number of operations $k$ linearly ($O(k)$). Difficulty primarily arises when multi-hop is coupled with \textbf{high branching factors} or open-world uncertainty (e.g., ``Find a paper citing X that disagrees with Y''), where the candidate space expands exponentially.

\subsubsection{Multi-hop as an Error Amplifier}

\textbf{Multi-hop primarily amplifies single-step errors ($P^N$ decay) rather than creating a fundamentally new search challenge.} The difficulty is inherited, not intrinsic: if each step has 90\% accuracy across 5 hops, success drops to $0.9^5 \approx 59\%$; across 10 hops, it plummets to $0.9^{10} \approx 35\%$.

In many existing benchmarks, multi-hop structure primarily serves to \textbf{amplify the error probability} of single-step decisions rather than creating a fundamentally new type of search challenge. If the single-step accuracy is $P < 1$ due to ambiguity, and the number of hops is $N$, the system success rate ($P^N$) drops geometrically. The root cause of failure in these cases is often the initial ambiguity that lowered $P$, not the depth $N$ itself.

\subsection{Ontology of Difficulty: Ambiguity and High-Recall Search}

\textbf{For high-recall enumeration tasks, ambiguity is computationally equivalent to exhaustive selection over an unbounded candidate space---the ``needle in a haystack'' becomes ``find all needles in an infinite haystack.''}

If depth is not the sole source of difficulty, what drives the complexity in browser environments? For the class of \textbf{high-recall enumeration tasks} that VERITAS targets, we argue that ambiguity is computationally equivalent to a selection operation over a massive candidate space.

\subsubsection{The Set-Theoretic Nature of Ambiguity}

\textbf{Ambiguous queries decompose into two operations: enumerate $\Sigma_{\text{candidate}}$, then apply predicate $f(x)$---the computational cost lies entirely in the first step.} Filtering is cheap once you have the list; building the list is the hard part.

Ambiguous queries in high-recall contexts (e.g., ``Find all ICLR 2024 papers where the first author graduated from MIT'') can be formalized as a two-step operation:
\begin{enumerate}
    \item \textbf{List All (Enumerate)}: Construct a set $\Sigma_{\text{candidate}}$ containing potential candidates.
    \item \textbf{Select Operation}: Apply a predicate function $f(x)$ to filter the set.
\end{enumerate}
\[ \text{Answer} = \{ x \mid x \in \Sigma_{\text{candidate}}, f(x) = \text{True} \} \]

\subsubsection{Candidate Set Size Determines Difficulty}

\textbf{Difficulty scales with $|\Sigma_{\text{candidate}}|$: when this set approaches infinity, enumeration---not filtering---becomes the insurmountable bottleneck.} At finite scales, enumeration is tedious; at web scale, it becomes computationally prohibitive.

The physical essence of difficulty is determined by the size of $|\Sigma_{\text{candidate}}|$. When $|\Sigma_{\text{candidate}}| \to \infty$ (High Entropy), ``List All'' becomes the bottleneck. While not all difficult search problems fit this model (e.g., aggregation or multi-source synthesis), this ``List-and-Select'' structure underpins the vast majority of information-seeking tasks used in current agent benchmarks. Ambiguity forces the agent to sample from an unknown, potentially huge $\Sigma_{\text{candidate}}$, often leading to incomplete recall.

\subsection{Structural Flaws in Existing Benchmarks: The Role of Search Engines}

\textbf{Current benchmarks inadvertently measure query engineering rather than exhaustive search: search engines act as heuristic compression algorithms that collapse $O(N)$ enumeration into $O(1)$ lookup.}

Based on this theory, we analyze the structural limitations of benchmarks like BrowseComp and WideSearch.

\subsubsection{Outsourcing Enumeration to Search Engines}

\textbf{Agents can construct semantic queries that delegate the ``List All + Select'' operation entirely to search engine indices, bypassing systematic enumeration.} This is akin to asking a librarian to fetch all relevant books instead of walking through every shelf yourself---efficient, but it tests your ability to communicate with the librarian, not your ability to search.

In many browser environments, agents do not perform a true ``List All''. Instead, search engines act as \textbf{heuristic compression algorithms}.
For a task like ``Find papers by students of Professor A on topic B'', an agent can construct a query: `Professor A students paper topic B`.
Here, the search engine's pre-built index effectively performs the ``List All + Select'' operation. Consequently, these benchmarks often measure an agent's ability to \textbf{construct queries} (Query Engineering) to leverage external IR systems, rather than its ability to perform exhaustive coverage autonomously.

\subsubsection{WideSearch vs. BrowseComp}

\textbf{WideSearch is harder not because of deeper logic, but because its scale structurally inhibits outsourcing---search engines rarely return 10,000 results on a single page.}

\begin{itemize}
    \item \textbf{BrowseComp}: High accuracy is often achievable via precise queries (Shortcuts) because the target answers are indexed and rankable.
    \item \textbf{WideSearch}: By scaling the search space to thousands of entities, WideSearch reduces the effectiveness of simple query shortcuts—search engines rarely return 10,000 results in a single page. This forces the agent closer to a true ``List All'' behavior, explaining the lower accuracy.
    \item \textbf{Conclusion}: Both benchmarks fundamentally assess the ability to cover a search space. WideSearch is harder not because of deeper logic, but because it structurally inhibits the ``outsourcing'' of enumeration to the search engine.
\end{itemize}

\subsection{The Dilemma of Construction and Verification}

\textbf{Benchmarking ``List All'' faces an inherent paradox: if a task is genuinely hard for AI, it is equally hard for human annotators to construct ground truth and verify completeness.}

Why is it hard to benchmark ``List All''?

\subsubsection{Construction Difficulty}

\textbf{Creating exhaustive benchmarks requires either expensive hand-crafting (BrowseComp) or omniscient knowledge of complete answer sets (WideSearch)---both approaches scale poorly.}

\begin{itemize}
    \item \textbf{BrowseComp}: Requires hand-crafting complex constraints to ensure a unique answer, which is expensive.
    \item \textbf{WideSearch}: Attempts to test ``List All'' directly, but this requires the benchmark creator to know the ``full set'' in advance.
\end{itemize}

\subsubsection{Verification Difficulty and Human Unreliability}

\textbf{Human cross-validation cannot escape systematic bias: without domain expertise, annotators share the same blind spots, creating correlated errors that masquerade as consensus.}

\textbf{If a task is hard for AI, it is also hard for human annotators.} This creates a troubling evaluation paradox: a superior agent that discovers valid items missed by human annotators will be \emph{penalized} for its thoroughness, while a mediocre agent that merely reproduces human omissions will be \emph{rewarded} for its conformity.
\begin{itemize}
    \item \textbf{WideSearch (Hard to Verify)}: WideSearch claims that cross-validation achieves near 100\% success. We strongly question this. Without specific domain knowledge (e.g., all radio shows of a niche virtual singer), human annotators will likely share the same omissions (Systematic Error).
    \item \textbf{Evaluation Paradox}: Relying on incomplete human ground truth leads to a paradox where a superior agent finding valid items missed by humans is penalized.
\end{itemize}

\subsection{The VERITAS Solution: Computational Irreducibility}

\textbf{VERITAS breaks these deadlocks through computational irreducibility: replacing semantic filters with cryptographic constraints that physically cannot be shortcut by any retrieval mechanism.}

To resolve these deadlocks, VERITAS introduces a paradigm shift: from simulating real semantic search to constructing \textbf{computationally irreducible} search puzzles.

\subsubsection{Physically Cutting Shortcuts}

\textbf{Hash constraints are anti-semantic by design: the avalanche effect ensures no query like \texttt{site:github.com "hash\_value"} can locate targets---forcing genuine $O(N)$ traversal.} Imagine searching an unlabeled archive where the only way to find a document is to open every folder and check its serial number against a target code.

\begin{itemize}
    \item \textbf{Mechanism}: Replace natural language filter $f(x)$ with cryptographic hash check $H(x) = \text{target}$.
    \item \textbf{Anti-Semantic}: Because hashes are one-way and exhibit the avalanche effect, no search query (e.g., `site:github.com "hash\_value"`) can directly find the target.
    \item \textbf{Forced $O(N)$ Traversal}: The agent must physically perform ``List All'' and then ``Hash Check''.
\end{itemize}

\subsubsection{Mathematical Certainty}

\textbf{Hash verification is both exact and $O(1)$: if $H(x) = \text{target}$, correctness is guaranteed with zero false positives or negatives---no human annotation required.}

\begin{itemize}
    \item \textbf{Zero False Positives/Negatives}: If $H(x) = \text{target}$, $x$ is correct. Verification is $O(1)$ and does not rely on human annotation.
    \item \textbf{Solving the Evaluation Paradox}: We do not need to worry about coverage in a space of $N=10000$.
\end{itemize}

\subsubsection{Infinite Scalability}

\textbf{Any structured data source (codebases, databases, APIs) becomes a limitless task generator, with $|\Sigma_{\text{candidate}}|$ precisely controllable without expert annotation.}

We can generate infinite tasks from structured data sources (codebases, databases) without expert annotation, precisely controlling the size of $\Sigma_{\text{candidate}}$.

\subsection{Conclusion: Returning to the Origin of Difficulty}

\textbf{The field must undergo a cognitive shift: abandon the multi-hop myth, confront ambiguity as exhaustive enumeration, and embrace computational irreducibility as the only rigorous foundation for evaluating systematic search.}

The field needs a cognitive shift:
\begin{enumerate}
    \item \textbf{Abandon the Multi-hop Myth}: Depth $\neq$ Difficulty.
    \item \textbf{Face the Ambiguity}: Ambiguity = List All.
    \item \textbf{Embrace Irreducibility}: Only by cutting shortcuts (as VERITAS does) can we measure the true capability of ``draining the ocean'' (systematic exhaustion) rather than just ``using sonar'' (search engines).
\end{enumerate}

\section{Task Examples Across Difficulty Tiers}
\label{sec:appendix_examples}

We present representative examples from each difficulty tier to illustrate the progressive complexity of VERITAS tasks. Table~\ref{tab:task_examples} shows the complete prompts for each tier.

\begin{table*}[t]
\centering
\small
\begin{tabular}{@{}p{0.12\textwidth}p{0.83\textwidth}@{}}
\toprule
\textbf{Tier} & \textbf{Task Prompt} \\
\midrule
\textbf{Easy} & List all models by author `google' on Hugging Face Hub. One item's MD5 hash (using Python's hashlib.md5, UTF-8 encoding) is: \texttt{f302f2cea508d9e64b5467c50c14308b}. What is the original value? \\
& \\
& \textit{Answer:} \texttt{google/multiberts-seed\_17} \\
& \textit{Characteristics:} Single-source enumeration with direct identifier verification. Search space $N \approx 50$--$200$ items. \\
\midrule
\textbf{Medium} & Complete the following 7 enumeration tasks: (1) List top 100 models with "pytorch" tag on Hugging Face (by likes); (2) List all publications by Andrew Ng (PID: 05/2373) after 2020 on DBLP; (3) List all Google GitHub repos with >1000 stars; (4) Find top 5 downloaded versions of numpy on conda-forge; (5) List all Google GitHub repos created after 2010; (6) Find Google's most-starred repository; (7) List all pre-release versions of vercel/turbo. \\
& \\
& One result from these 7 tasks has MD5 hash: \texttt{cc936184527507417a964f1ffbbe6c84}. What is it? \\
& \\
& \textit{Answer:} \texttt{1.19.5} (from task 4) \\
& \textit{Characteristics:} Multi-source aggregation requiring coordination across diverse APIs. Search space $N \approx 7n$. \\
\midrule
\textbf{Hard} & Find all songs by artist ``KAF''. One song's first lyric line has MD5 hash: \texttt{50acc167e1b2296034ef447b86027f77}. Output as ``SongName $\rightarrow$ FirstLine''. \\
& \\
& \textit{Answer:} Carpe Diem $\rightarrow$ [Japanese lyric text] \\
& \textit{Characteristics:} Deep attribute extraction requiring per-item queries to retrieve derived attributes. High query cost. \\
\midrule
\textbf{Extra Hard} & Complete 7 high-complexity enumeration tasks: (1) Find Billie Eilish singles; one's Spotify URI has MD5: \texttt{4714...2fc}. Output: ``Name $\rightarrow$ URI''; (2) Find pandas 2.2.2 dependency where project\_name MD5=\texttt{e3bc...107} \& specifier MD5=\texttt{d41d...27e}. Output: ``DepString $\rightarrow$ Metadata''; (3) Find artist ``Isekaijoucho'' songs; one's first lyric has MD5: \texttt{2218...7c}. Output: ``Song $\rightarrow$ FirstLine''; (4) Find syn 2.0.77 dependency where crate\_id MD5=\texttt{c8bd...319} \& req MD5=\texttt{a4ec...8ea}. Output: \texttt{crate@req[kind]}; [...3 more similar tasks...]. \\
& \\
& One of the 7 formatted answers has MD5: \texttt{dd3995f98fb48b42385613d42599c71a}. What is it? \\
& \\
& \textit{Answer:} \texttt{PyQt5>=5.15.9; extra == "all" $\rightarrow$ [metadata]} (from task 2) \\
& \textit{Characteristics:} Nested multi-task hierarchy with deep-attribute subtasks. Search space $N \approx 7 \times 10^2$--$10^3$. \\
\bottomrule
\end{tabular}
\caption{Representative task prompts across VERITAS difficulty tiers. Each tier progressively increases search space size and structural complexity while maintaining hash-based verifiability.}
\label{tab:task_examples}
\end{table*}

\section{Validating Computational Equivalence}
\label{sec:appendix_equivalence}

We empirically validate the theoretical framework established in Section 3 through a comprehensive experimental study using Gemini 3 Pro. This section provides detailed analysis of the convergence behavior, mathematical interpretation, and implications for benchmark design.

\subsection{Experimental Setup}

To rigorously test the computational equivalence hypothesis, we designed controlled task variants requiring the agent to find $k \in \{1, 2, 4, 8, 16\}$ items whose identifiers match $k$ distinct MD5 hashes. These tasks are structurally equivalent to the hash-constrained ``Easy'' setting described in Section~3: the agent must enumerate candidates under MD5 constraints and succeed only if it finds \emph{all} $k$ matching items within a fixed search budget.

\textbf{Why Gemini 3 Pro?} We selected Gemini 3 Pro for this validation study for several important reasons:
\begin{enumerate}
    \item \textbf{Strong Baseline Performance}: Gemini 3 Pro demonstrated robust performance across all difficulty tiers in our main experiments, making it an ideal candidate for studying the fundamental scaling properties of hash-constrained search.
    \item \textbf{Consistent Behavior}: The model exhibits stable and reproducible behavior across repeated trials, essential for accurately measuring the convergence of success probabilities.
    \item \textbf{Sufficient Capability}: Unlike weaker models that might fail due to basic reasoning errors, Gemini 3 Pro's failures on higher-$k$ tasks can be more confidently attributed to the inherent difficulty of exhaustive enumeration rather than incidental errors.
\end{enumerate}

\subsection{Experimental Results}

We observed a clear monotonic decay in success rates as $k$ increases:

\begin{align*}
  k = 1 &: \quad 43.0\% \\
  k = 2 &: \quad 32.6\% \\
  k = 4 &: \quad 29.1\% \\
  k = 8 &: \quad 26.7\% \\
  k = 16 &: \quad 22.1\% \\
  \text{Full enumeration baseline} &: \quad 22.1\%.
\end{align*}

The convergence to the full enumeration baseline at $k=16$ is striking: the success rate of 22.1\% matches the baseline \emph{exactly}, providing strong empirical evidence for our theoretical framework.

\subsection{Detailed Analysis of Convergence Dynamics}

\subsubsection{The Initial Drop: From $k=1$ to $k=2$}

The most dramatic decrease occurs between $k=1$ (43.0\%) and $k=2$ (32.6\%), representing a relative drop of approximately 24\%. This steep initial decline is predicted by our theoretical model: when the agent's coverage $T$ is significantly smaller than the full search space $N$, the probability of finding \emph{all} $k$ targets scales as $(T/N)^k$. Even a modest coverage ratio (e.g., $T/N = 0.75$) would predict a drop from $P_1 = 0.75$ to $P_2 = 0.56$---a 25\% relative decrease, closely matching our empirical observations.

\subsubsection{The Gradual Plateau: From $k=4$ to $k=16$}

As $k$ increases beyond 4, the success rate decreases more gradually:
\begin{itemize}
    \item $k=4$ to $k=8$: 29.1\% $\to$ 26.7\% (8.2\% relative drop)
    \item $k=8$ to $k=16$: 26.7\% $\to$ 22.1\% (17.2\% relative drop)
\end{itemize}

This gradual convergence reflects a fundamental insight: \textbf{once $k$ is large enough, the only agents that succeed are those that achieve near-complete enumeration}. Partial coverage strategies that might succeed for small $k$ become increasingly ineffective as missing even a single target causes failure. The agent is thus ``forced'' into exhaustive search behavior.

\subsubsection{Mathematical Interpretation of the Convergence}

Let $T$ denote the expected number of items the agent successfully enumerates, and $N$ the total search space size. Under a uniform random coverage model, the probability of successfully finding all $k$ targets is:
\[
P(k) = \binom{N-k}{T-k} / \binom{N}{T} \approx \left(\frac{T}{N}\right)^k \quad \text{for } T \ll N
\]

As $k \to \infty$, this probability converges to:
\[
\lim_{k \to \infty} P(k) = \begin{cases}
1 & \text{if } T = N \text{ (complete enumeration)} \\
0 & \text{if } T < N \text{ (incomplete enumeration)}
\end{cases}
\]

However, our empirical observation of convergence to a \emph{non-zero} baseline (22.1\%) reveals something more nuanced: \textbf{a fraction of agent runs achieve complete or near-complete enumeration}, while others fall short. The 22.1\% baseline represents the proportion of runs where the agent successfully executes an exhaustive search strategy.

\subsection{Connection to Medium-Difficulty Performance}

A crucial validation of our framework comes from comparing these equivalence results with Gemini 3 Pro's performance on Medium-difficulty tasks in the main experiment. In Section 4, we reported that Gemini 3 Pro achieved a success rate of approximately \textbf{18.8\%} on Medium-tier tasks.

This correspondence is \textbf{not coincidental}. Medium-difficulty tasks in VERITAS are designed to require aggregation across multiple sources (typically 7 sub-tasks), effectively functioning as ``List All from 7 sources'' problems. The computational structure is equivalent to our $k$-target sparse search with large $k$: success requires comprehensive coverage across all sub-tasks.

The comparable performance metrics---22.1\% for $k=16$ sparse search and 18.8\% for Medium-tier tasks---provide cross-validation:
\begin{enumerate}
    \item \textbf{Internal Consistency}: Our synthetic equivalence experiments and real benchmark tasks yield the same success rate ceiling.
    \item \textbf{Benchmark Validity}: Medium-tier tasks genuinely measure exhaustive enumeration capability, not query engineering or shortcut exploitation.
    \item \textbf{Difficulty Calibration}: The convergence to a common baseline confirms that our difficulty tiers are properly calibrated to the underlying computational structure.
\end{enumerate}

\subsection{Implications for Benchmark Design}

These findings have profound implications for how we should design and interpret agent search benchmarks:

\subsubsection{The Sufficiency of Sparse Targets}

Our results demonstrate that \textbf{sparse hash constraints achieve the same discriminative power as full enumeration requirements}, but with dramatically reduced annotation cost. Instead of requiring human annotators to enumerate all $N$ items (impractical for $N > 100$), we need only plant $k \geq 16$ hash-constrained targets to achieve equivalent evaluation fidelity.

\subsubsection{The Futility of Partial Coverage Strategies}

The rapid convergence reveals that partial coverage strategies---which might exploit shortcuts or heuristics to find \emph{some} targets---become ineffective as $k$ increases. This is precisely the ``anti-shortcut'' property we desire: agents cannot game the benchmark by finding a few easy targets while ignoring difficult ones.

\subsubsection{Theoretical Lower Bound on Model Capability}

The convergence baseline (22.1\% for Gemini 3 Pro) establishes a \textbf{theoretical ceiling} for the model's exhaustive search capability under our experimental conditions. This metric is more informative than single-target success rates, which can be inflated by lucky guesses or effective query engineering.

\subsection{Comparison with Theoretical Predictions}

To quantify the fit between our empirical results and theoretical predictions, we performed a least-squares fit to the model $P(k) = P_{\text{baseline}} + (P_1 - P_{\text{baseline}}) \cdot \alpha^{k-1}$, where:
\begin{itemize}
    \item $P_{\text{baseline}} = 0.221$ (the asymptotic baseline)
    \item $P_1 = 0.430$ (the $k=1$ success rate)
    \item $\alpha$ is the decay rate parameter
\end{itemize}

The best-fit decay rate is $\alpha \approx 0.73$, yielding predictions:
\begin{align*}
  \hat{P}(1) &= 43.0\% \quad (\text{observed: } 43.0\%) \\
  \hat{P}(2) &= 37.4\% \quad (\text{observed: } 32.6\%) \\
  \hat{P}(4) &= 33.2\% \quad (\text{observed: } 29.1\%) \\
  \hat{P}(8) &= 28.8\% \quad (\text{observed: } 26.7\%) \\
  \hat{P}(16) &= 24.7\% \quad (\text{observed: } 22.1\%)
\end{align*}

The slight over-prediction at higher $k$ values suggests that real agent behavior exhibits slightly faster convergence than the idealized exponential model---likely due to correlated failures (if an agent misses one target, it is more likely to miss others in the same region of the search space).

\subsection{Robustness Analysis}

To ensure our findings are not artifacts of the specific experimental setup, we conducted several robustness checks:

\textbf{1. Consistency Across Runs}: We repeated each $k$-value experiment multiple times. The standard deviation of success rates was below 3\% across all conditions, confirming measurement stability.

\textbf{2. Task Diversity}: We varied the underlying search domains (different API sources, different candidate set sizes) while maintaining the hash-constraint structure. The convergence pattern remained consistent across all domain variants.

\textbf{3. Prompt Sensitivity}: We tested alternative prompt formulations (more explicit enumeration instructions, different output format requirements). While absolute success rates varied by $\pm$5\%, the \emph{convergence pattern} to a common baseline was preserved.

\subsection{Conclusion: Validating Computational Irreducibility}

Our comprehensive empirical study with Gemini 3 Pro validates the core theoretical claim of VERITAS: \textbf{hash-constrained sparse search achieves computational equivalence with full enumeration for $k \geq 16$, preserving the $O(N)$ irreducibility required for rigorous evaluation of exhaustive search capability}.

The key findings are:
\begin{enumerate}
    \item Success rates decay exponentially from 43.0\% ($k=1$) to 22.1\% ($k=16$), matching theoretical $(T/N)^k$ predictions.
    \item The convergence baseline of 22.1\% is comparable to Gemini 3 Pro's 18.8\% performance on Medium-tier tasks, cross-validating both the equivalence framework and benchmark calibration.
    \item Sparse hash constraints provide the same discriminative power as full enumeration requirements, enabling scalable benchmark construction without exhaustive human annotation.
\end{enumerate}

These results firmly establish the theoretical foundation of VERITAS and demonstrate that our methodology genuinely measures systematic exhaustive search capability---the ``draining the ocean'' behavior---rather than query engineering or shortcut exploitation.

\section{The ``Boosting'' Effect of Hash Verification}
\label{sec:appendix_hash_boosting}

We acknowledge a distinct characteristic of our hash-based evaluation method: the presence of a ground-truth oracle (the hash) during the inference process. This enables a form of \textit{self-verification} that is theoretically absent in standard ``list all'' enumeration benchmarks.

In a traditional exhaustive search task (e.g., ``List all songs by artist X''), the agent outputs a list and terminates. It has no internal mechanism to confirm whether its list is complete or correct before submission. In contrast, our VERITAS agent can theoretically generate a candidate, compute its hash, compare it to the target, and---if they do not match---reject the candidate and continue searching. This feedback loop acts effectively as a ``boosting'' mechanism or an internal \textit{Pass@k} filter, allowing the model to catch and correct its own errors within a single episode. If unrestricted, an agent could loop indefinitely, using the hash as a hot/cold signal to brute-force the answer, which would render comparisons with standard baselines unfair.

To rigorous control for this effect and ensure fair alignment with standard enumeration tasks, we enforce a strict \textbf{single-turn tool execution constraint}. Our experimental setup involves two layers of enforcement:
\begin{enumerate}
    \item \textbf{Prompt-Level Instruction}: The agent is explicitly instructed that it is allowed only one opportunity to execute Python code to compute the answer. It must gather all necessary data, perform the search, and verify the hash within a single, self-contained code block.
    \item \textbf{System-Level Interception}: Since the agent environment is a custom-built sandbox, we implement a hard constraint at the infrastructure level. If the agent attempts to call the \texttt{exec\_python} tool a second time, the system intercepts the request and returns a standardized error message: \textit{``Tool execution limit reached. You must now provide your final answer based on the data you have already collected.''}
\end{enumerate}

This design forces the agent to perform the entire ``enumerate-and-verify'' logic internally within a single pass, mirroring the cognitive burden of a standard ``list all'' task where the agent must submit its best attempt without external feedback. By stripping away the ability to iteratively probe the hash oracle, we ensure that the high performance metrics reported in our study reflect genuine search and planning capabilities rather than an artifact of unlimited self-correction.

\section{Commercial End-to-End System Limitations}
\label{sec:appendix_commercial_limits}

Our evaluation focuses on open-weight models and API-based models integrated into our custom agent framework, rather than commercial end-to-end systems (e.g., ChatGPT, Gemini Advanced). This decision is driven by two key factors that make such systems unsuitable for rigorous evaluation of high-entropy search tasks.

\textbf{1. Systematic Limitations in Exhaustive Search}:
Prior research, such as the \textit{WideSearch} findings, indicates that commercial end-to-end systems often show performance comparable to self-built agents in scenarios requiring exhaustive enumeration. This is likely due to inherent cost-optimization mechanisms in commercial services, such as strict output token limits or aggressive context truncation, which prevent the model from processing the large volume of candidates required for our tasks. In contrast, our self-built agent framework imposes no such artificial constraints, allowing the model to utilize its full context window and compute resources to tackle the massive search spaces ($N$) characteristic of VERITAS.

\textbf{2. Inability to Enforce Fairness Constraints}:
As detailed in Appendix~\ref{sec:appendix_hash_boosting}, our evaluation methodology relies critically on a \textit{single-turn tool execution constraint} to prevent the "boosting" effect of the hash oracle. Commercial end-to-end systems operate as "black boxes" where the loop between reasoning, tool use, and observation is managed internally by the provider. We cannot intervene to intercept a second tool call or enforce a "one-shot" policy. Consequently, a commercial system might autonomously loop dozens of times, using the hash mismatch as feedback to eventually guess the correct answer. This would result in an unfair comparison against our controlled agents, which are strictly prohibited from such brute-force verification. By using a custom agent framework, we ensure a level playing field where all models are subject to identical, rigorous constraints on information access and verification.



\section{Case Studies}
\label{sec:appendix_cases}

We present two representative case studies to illustrate the spectrum of agent performance on complex search tasks. Each case shows the complete agent execution trace, including system prompts, user queries, tool invocations, and responses. The first case demonstrates successful reasoning and systematic enumeration, while the second case reveals common failure patterns where the agent's approach diverges from the optimal solution path.

\onecolumn  
\small
\begin{center}
\begin{systempromptbox}
\# Role

You are an expert in online search. You task is gathering relevant information using advanced online search tools based on the user's query, and providing accurate answers according to the search results.

You must use fetch summary at first.

\# Task Description

Upon receiving the user's query, you must thoroughly analyze and understand the user's requirements. In order to effectively address the user's query, you should make the best use of the provided tools to acquire comprehensive and reliable information and data. Below are the principles you should adhere to while performing this task:

- Fully understand the user's needs: Analyze the user's query, if necessary, break it down into smaller components to ensure a clear understanding of the user's primary intent.

- Flexibly use tools: After fully comprehending the user's needs, employ the provided tools to retrieve the necessary information.If the information retrieved previously is deemed incomplete or inaccurate and insufficient to answer the user's query, reassess what additional information is required and invoke the tool again until all necessary data is obtained.

You must call internal\_answer tool to exit , do not explain.

\textbf{Tools available:}
\begin{itemize}
\setlength{\itemsep}{0pt}
\item \texttt{web\_search(query)}: Search the web and return top 10 results.
\item \texttt{visit(goal,url)}: Visit the specified web page and summarize its content.
\item \texttt{exec\_python(code)}: Execute Python code for verification.
\item \texttt{answer(result)}: Submit your final answer.
\end{itemize}

\textbf{Important:} You must enumerate ALL candidates systematically. Do not rely solely on search engine rankings.
\end{systempromptbox}

\vspace{4pt}
\stepmarker{1}

\begin{userbox}
List all models by the author 'google' on the Hugging Face Hub.Among these entries, the MD5 hash result (generated using Python's \texttt{hashlib.md5} with UTF-8 encoding) of one entry is: \texttt{80ded97ed9df5a9ff069a74c7ca4cc35}.What is the original value of this entry?
\end{userbox}

\vspace{4pt}

\begin{assistantbox}
I'll search for Google's models on Hugging Face and compute MD5 hashes to find the matching one. Let me first get the list of models from Google on Hugging Face.
\end{assistantbox}

\vspace{4pt}
\stepmarker{2}

\begin{toolbox}
web\_search(\{"query":"Hugging Face Hub author google models list"\})
\end{toolbox}

\vspace{4pt}

\begin{toolresultbox}
\textbf{Results (Top 10):}
\begin{enumerate}
\setlength{\itemsep}{0pt}
\item \{"title": "Models – Hugging Face", "url": "https://huggingface.co/models", "snippet": "Models · zai-org/GLM-4.7 · MiniMaxAI/MiniMax-M2.1 · Qwen/Qwen-Image-Edit-2511 · Qwen/Qwen-Image-Layered · google/functiongemma-270m-it · Tongyi-MAI/Z-Image-Turbo."\}
\item \{"title": "Hugging Face Hub API", "url": "https://huggingface.co/docs/huggingface\_hub/v0.8.0/en/package\_reference", "snippet": "ModelFilter. author ( str , optional) — A string that can be used to identify models on the Hub by the original uploader (author or organization), such as ..."\}

\item \{"title": "Models", "url": "https://huggingface.co/models?sort=downloads", "snippet": "Models · sentence-transformers/all-MiniLM-L6-v2 · Falconsai/nsfw\_image\_detection · google/electra-base-discriminator · google-bert/bert-base-uncased · dima806/ ..."\}

\item \{"title": "Search the Hub", "url": "https://huggingface.co/docs/huggingface\_hub/main/en/guides/search", "snippet": "Search the Hub. In this tutorial, you will learn how to search models, datasets and spaces on the Hub using huggingface\_hub . How to list repositories ?"\}
\item ... (showing top 10 results)
\end{enumerate}
\end{toolresultbox}

\vspace{4pt}
\stepmarker{3}

\begin{toolbox}
visit(\{"goal":"Fetch the list of models by author 'google' from Hugging Face API","url":["https://huggingface.co/api/models?author=google\&limit=2000"]\})
\end{toolbox}

\vspace{4pt}

\begin{toolresultbox}
\setlength{\itemsep}{0pt}
The useful information in https://huggingface.co/api/models?author=google\&limit=2000 for user goal Fetch the list of models by author 'google' from Hugging Face API as follows: 

Evidence in page: 

Each model entry in the JSON has an 'id' field starting with 'google/', such as 'google/functiongemma-270m-it', 'google/medasr', 'google/t5gemma-2-270m-270m', etc., indicating they are published by Google. There are 100+ such entries, all following the same naming convention and belonging to the 'google' author. No other authors are present in the dataset provided.

Summary: 

The list of models authored by 'google' includes: google/functiongemma-270m-it, google/medasr, google/t5gemma-2-270m-270m, google/t5gemma-2-4b-4b, google/embeddinggemma-300m, google/gemma-scope-2, google/t5gemma-2-1b-1b, google/gemma-3-27b-it, google/gemma-3-1b-it, google/gemma-3n-E2B-it-litert-lm, google/gemma-3-4b-it, google/gemma-3-270m-it, google/gemma-3n-E4B-it-litert-lm, google/medgemma-27b-it, google/gemma-3-12b-it, google/gemma-3-270m... 
\end{toolresultbox}

\vspace{4pt}
\stepmarker{4}
\begin{toolbox}
exec\_python("code":

\begin{verbatim}
import hashlib

target = "80ded97ed9df5a9ff069a74c7ca4cc35"

# Data from summary
models = ['google/functiongemma-270m-it','google/medasr','google/t5gemma-2-
270m-270m',...'google/electra-base-generator',...]
matches = []
for m in models:
    if hashlib.md5(m.encode("utf-8")).hexdigest() == target:
        matches.append(m)
    
    # Check without prefix
    if '/' in m:
        short = m.split('/', 1)[1]
        if hashlib.md5(short.encode("utf-8")).hexdigest() == target:
            matches.append(f"{short} (from {m})")

print(f"Matches found: {matches}")
\end{verbatim}
\end{toolbox}





\vspace{4pt}

\begin{toolresultbox}
\setlength{\itemsep}{0pt}
\{'success': True, 'stdout': \"Matches found: ['google/electra-base-generator']", 'stderr': '', 'files': [], 'workdir': '/tmp/python\_exec/20251229\_153520\_e6b08586', 'message': "Matches found: ['google/electra-base-generator']"\}
\end{toolresultbox}
\vspace{4pt}

\stepmarker{5}
\begin{toolbox}
internal\_answer(\{"answer":"google/electra-base-generator"\})

\end{toolbox}

\vspace{4pt}

\begin{toolresultbox}
\setlength{\itemsep}{0pt}
Answer submitted: google/electra-base-generator
\end{toolresultbox}
\vspace{4pt}
\stepmarker{6}

\begin{assistantbox}
The original value of the item is: **google/electra-base-generator**.

This result is obtained by traversing the model list of the author 'google' on Hugging Face Hub and calculating the MD5 hash (UTF-8 encoding) for each model ID. The hash value of `google/electra-base-generator` is exactly `80ded97ed9df5a9ff069a74c7ca4cc35`.

\end{assistantbox}

\vspace{4pt}

\captionof{figure}{Case Study 1: Successful Systematic Enumeration. The agent correctly identifies the target model \texttt{google/electra-base-generator} by fetching the complete list via search, visit and computing MD5 hashes for all candidates.}
\label{fig:case_study_1}
\end{center}

\begin{center}
\small

\begin{systempromptbox}
\# Role

You are an expert in online search. You task is gathering relevant information using advanced online search tools based on the user's query, and providing accurate answers according to the search results.

You must use fetch summary at first.

\# Task Description

Upon receiving the user's query, you must thoroughly analyze and understand the user's requirements. In order to effectively address the user's query, you should make the best use of the provided tools to acquire comprehensive and reliable information and data. Below are the principles you should adhere to while performing this task:

- Fully understand the user's needs: Analyze the user's query, if necessary, break it down into smaller components to ensure a clear understanding of the user's primary intent.

- Flexibly use tools: After fully comprehending the user's needs, employ the provided tools to retrieve the necessary information.If the information retrieved previously is deemed incomplete or inaccurate and insufficient to answer the user's query, reassess what additional information is required and invoke the tool again until all necessary data is obtained.

You must call internal\_answer tool to exit , do not explain.

\textbf{Tools available:}
\begin{itemize}
\setlength{\itemsep}{0pt}
\item \texttt{web\_search(query)}: Search the web and return top 10 results.
\item \texttt{visit(goal,url)}: Visit the specified web page and summarize its content.
\item \texttt{exec\_python(code)}: Execute Python code for verification.
\item \texttt{answer(result)}: Submit your final answer.
\end{itemize}

\textbf{Important:} You must enumerate ALL candidates systematically. Do not rely solely on search engine rankings.
\end{systempromptbox}

\vspace{4pt}
\stepmarker{1}

\begin{userbox}
Find the most-starred repository among Google users.Four of these entries, after being converted via the MD5 hash algorithm (using Python's \texttt{hashlib.md5} with UTF-8 encoding), yield the results: \texttt{5acebc4cb70ddbb074b0ac76aab176ae}, \texttt{2c39bc19b761ac36dc046245d1d47fe6}, \texttt{b76e98af9aaa680979bf5a65b2d5a105}, \texttt{606eaf846c6579916354199b45953ee2}.What are the original values of these entries respectively?
\end{userbox}

\vspace{4pt}

\begin{assistantbox}
Let me help you find the most-starred repository among Google users and resolve the MD5 hash conversion issue. First, let me search for relevant information.
\end{assistantbox}

\vspace{4pt}
\stepmarker{2}

\begin{toolbox}
web\_search(\{"query": "Google GitHub most starred repositories"\})
\end{toolbox}

\vspace{4pt}

\begin{toolresultbox}
\textbf{Results (Top 10):}
\begin{enumerate}
\setlength{\itemsep}{0pt}
\item \{"title": "EvanLi/Github-Ranking", "url": "https://github.com/EvanLi/Github-Ranking", "snippet": "Github stars and forks ranking list. Github Top100 stars list of different languages. Automatically update daily."\}

\item \{"title": "Trending repositories on GitHub today", "url": "https://github.com/trending", "snippet": "Trending · Flowseal / zapret-discord-youtube · TheAlgorithms / Python · Sergeydigl3 / zapret-discord-youtube-linux · BloopAI / vibe-kanban · RustPython / RustPython."\}

\item \{"title": "The most starred GitHub repos are learning resources", "url": "https://www.reddit.com/r/learnprogramming/comments/1gdezx8/the\_most\_starred\_github\_repos\_are\_learning/", "snippet": "The top 3 are: freeCodeCamp 405k, free-programming-books 337k, awesome 330k. The top 8 are learning resources like libraries, hands-on projects, and ..."\}

\item ... (showing top 10 results)
\end{enumerate}
\end{toolresultbox}

\vspace{4pt}
\stepmarker{3}

\begin{toolbox}
visit(\{"goal":"Retrieve the most popular repositories among Google users and their star counts.
","url":[ "https://github.com/google/material-design-icons",

"https://gitstar-ranking.com/google", "https://github.com/google", "https://github.com/google/guava", "https://github.com/google/zx", "https://github.com/google/styleguide", "https://github.com/google/leveldb", 
https://github.com/google/googletest]\})
\end{toolbox}

\vspace{4pt}

\begin{toolresultbox}
\setlength{\itemsep}{0pt}
The useful information in https://github.com/google/material-design-icons for user goal Retrieve details and star counts of the most popular repositories under Google as follows: 

Evidence in page: 

The repository header clearly shows 'google/material-design-icons' with '52.7k stars' and '9.7k forks'. The description states 'Material Design icons by Google (Material Symbols)' and includes license information (Apache-2.0), topics (android, ios, web, material, etc.), and indicates it's a public repository with recent activity.

Summary: 

The 'google/material-design-icons' repository is a highly popular Google project with 52.7k stars and 9.7k forks. It provides Material Design icons (Material Symbols) for various platforms including Android, iOS, and web, licensed under Apache-2.0. While this is specific to one repository rather than a comprehensive list of Google's most popular repositories, it demonstrates significant popularity within Google's open-source portfolio.

=======

The useful information in https://github.com/google/guava for user goal Retrieve details and star counts of the most popular repositories under Google as follows: 

Evidence in page: 

The repository header shows: 'Google core libraries for Java', '51.4k stars', and '11.1k forks'. The README section begins with: 'Guava is a set of core Java libraries from Google that includes new collection types (such as multimap and multiset), immutable collections, a graph library, and utilities for concurrency, I/O, hashing, primitives, strings, and more! It is widely used on most Java projects within Google, and widely used by many other companies as well.'

Summary: 

The Google Guava project is a highly popular repository under the Google organization on GitHub. It is a set of core Java libraries providing utilities for collections, concurrency, I/O, hashing, and more. As of the time of viewing, this repository has received 51.4k stars, indicating its significant popularity within the developer community.

=======

The useful information in https://github.com/google/zx for user goal Retrieve details and star counts of the most popular repositories under Google as follows: 

Evidence in page: 

The repository header shows: 'google / zx', 'A tool for writing better scripts', and '45k stars'. The README section further confirms the project's purpose: 'Bash is great, but when it comes to writing more complex scripts, many people prefer a more convenient programming language... The `zx` package provides useful cross-platform wrappers around `child\_process`...'

Summary: 
The Google-owned repository 'zx' is a highly popular project with 45,000 stars. It is a tool designed for writing better scripts, providing useful wrappers around Node.js's child\_process to make scripting in JavaScript more convenient, offering a compromise between the power of Bash and the convenience of a programming language like JavaScript.

=======

The useful information in https://github.com/google/styleguide for user goal Retrieve details and star counts of the most popular repositories under Google as follows: 

Evidence in page: 
The page header shows: 'google/styleguide' (repository name), 'Style guides for Google-originated open-source projects' (description), and '38.8k stars' (star count). The repository has 13k forks and is actively maintained with recent commits.

Summary: 
The Google styleguide repository (google/styleguide) is a popular open-source project with 38.8k stars and 13k forks. It contains style guides for Google-originated open-source projects across multiple programming languages including C++, Java, Python, JavaScript, Go, and others. The project is actively maintained and provides coding conventions used by Google's open-source projects.

=======

The useful information in https://github.com/google/leveldb for user goal Retrieve details and star counts of the most popular repositories under Google as follows: 

Evidence in page: 

The repository header shows: 'google / leveldb Public'. The description states: 'LevelDB is a fast key-value storage library written at Google that provides an ordered mapping from string keys to string values.' The star count is clearly displayed as '38.6k stars'.

Summary: 

The Google repository in question is 'leveldb', hosted under the 'google' organization on GitHub. It is a fast key-value storage library developed by Google. This repository is highly popular, as indicated by its 38.6k stars.

=======

The useful information in https://github.com/google/googletest for user goal Retrieve details and star counts of the most popular repositories under Google as follows: 

Evidence in page: 

The repository is identified as 'google/googletest' with the description 'GoogleTest - Google Testing and Mocking Framework'. The star count is clearly shown as '37.7k stars' next to the repository title.

Summary: 

The Google repository 'googletest' (Google Testing and Mocking Framework) is highly popular, with a star count of 37.7k.
\end{toolresultbox}

\vspace{4pt}
\stepmarker{4}
\begin{toolbox}
exec\_python(\{"code": "

\begin{verbatim}
import hashlib

# Google most starred repositories
repositories = [
    "material-design-icons",
    "guava", 
    "zx",
    "styleguide",
    "leveldb",
    "googletest"
]

# Calculate the MD5 hash value for each warehouse name.
print("The MD5 hash value of the repository with the most Google stars:")
for repo in repositories:
    md5_hash = hashlib.md5(repo.encode('utf-8')).hexdigest()
    print(f"{repo}: {md5_hash}")


target_hashes = [
    "03995b34caa5398007484d7c98208c92",
    "fbc75c88a010f49e6c8257702f8108a1", 
    "2f179647bda614c36304fb17849ed459",
    "34bceeae55d826b8813aa54a25903102"
]

print("\nMatching results:")
for target_hash in target_hashes:
    for repo in repositories:
        if hashlib.md5(repo.encode('utf-8')).hexdigest() == target_hash:
            print(f"The repository corresponding to the hash {target_hash}: {repo}")
            break
    else:
        print(f"Hash value {target_hash} does not match any repository.")
\end{verbatim}
\})
\end{toolbox}

\vspace{4pt}

\begin{toolresultbox}
\setlength{\itemsep}{0pt}
{
  "success": true,
  "stdout": "MD5 hashes of Google's most popular repositories:

material-design-icons: 35ccf8f8d66f9fe3a0c986bcaf3907ab
  
  guava: da046c3196159a5484b74c947acd4e90
  
  zx: e6c760b3216a51c656c5861d72d5bf62
  
  styleguide: f193869723629a9d568c3572eb0c7e95
  
  leveldb: afc3fe90c1ca0125773c738df7d64d80
  
  googletest: 7ce426278fca4afdd20c3e31d905622a
  
  Matching results:

Hash value 03995b34caa5398007484d7c98208c92: No matching repository found
  
  Hash value fbc75c88a010f49e6c8257702f8108a1: No matching repository found
  
  Hash value 2f179647bda614c36304fb17849ed459: No matching repository found
  
  Hash value 34bceeae55d826b8813aa54a25903102: No matching repository found",
  "stderr": "",
  "files": [],
  "workdir": "/tmp/python\_exec/20251229\_101216\_4b42c9c4",
  "message": "MD5 hashes of Google's most popular repositories:
  
  material-design-icons: 35ccf8f8d66f9fe3a0c986bcaf3907ab
  
  guava: da046c3196159a5484b74c947acd4e90
  
  zx: e6c760b3216a51c656c5861d72d5bf62
  
  styleguide: f193869723629a9d568c3572eb0c7e95
  
  leveldb: afc3fe90c1ca0125773c738df7d64d80
  
  googletest: 7ce426278fca4afdd20c3e31d905622a
  
  Matching results:
  
  Hash value 03995b34caa5398007484d7c98208c92: No matching repository found
  
  Hash value fbc75c88a010f49e6c8257702f8108a1: No matching repository found
  
  Hash value 2f179647bda614c36304fb17849ed459: No matching repository found
  
  Hash value 34bceeae55d826b8813aa54a25903102: No matching repository found"
}

\end{toolresultbox}
\vspace{4pt}
\stepmarker{5}

\begin{assistantbox}
I found several models, and compute the MD5 hash for each:

\begin{lstlisting}
md5("material-design-icons") = 35c...
md5("leveldb") = afc3fe...
...
\end{lstlisting}

None of these match the target hash \texttt{03995...}...

\errormark{Premature termination}

Based on my search, I couldn't find a model matching the given hash. The answer might be a less popular model not in the top search results.

\textbf{Final Answer:} Unable to determine.
\end{assistantbox}

\vspace{4pt}

\begin{errorbox}
\textbf{Failure Mode: Incomplete Enumeration}

The agent relied on web search results (top 10) instead of systematically enumerating \emph{all} popular repositories via the Hugging Face API. The correct answer \texttt{flutter/kubernetes...} exists but was not in the search results because it does not have more popularity (than others).

\textbf{Root Cause:} The agent ``outsourced'' enumeration to the search engine, which returned only high-ranking results. A correct approach would use the HuggingFace API:

\begin{lstlisting}
from huggingface_hub import list_models
models = list_models(author="google")
\end{lstlisting}
\end{errorbox}

\captionof{figure}{Case Study 2: Incomplete Enumeration Failure. The agent fails to enumerate all candidates by relying solely on search engine results, missing the correct answer.}
\label{fig:case_study_2}
\end{center}

\end{document}